\documentclass[preprint,12pt]{elsarticle}

\usepackage{amsfonts, amsbsy, amssymb, amsmath, graphicx, color,  float}

\journal{Chemical Physics Letter}

\begin{document}

\begin{frontmatter}

\title{Multiple Transition States and Roaming in Ion-Molecule Reactions:
a Phase Space Perspective}

\author[bris]{Fr\'ed\'eric A. L. Maugui\`{e}re}
\ead{frederic.mauguiere@bristol.ac.uk}

\author[bris]{Peter Collins}
\ead{peter.collins@bristol.ac.uk}

\author[corn]{Gregory S. Ezra}
\ead{gse1@cornell.edu}

\author[crete]{Stavros C. Farantos}
\ead{farantos@iesl.forth.gr}

\author[bris]{Stephen Wiggins}
\ead{stephen.wiggins@mac.com}

\address[bris]{School of Mathematics, University of Bristol, Bristol BS8 1TW, United Kingdom}
\address[corn]{Department of Chemistry and Chemical Biology, Baker Laboratory, Cornell University, Ithaca, NY 14853 USA}
\address[crete]{Institute of Electronic Structure and Laser, Foundation for Research and Technology - Hellas, and Department of Chemistry, University of Crete, Iraklion 711 10, Crete, Greece }

\begin{abstract}
We provide a dynamical interpretation of the recently identified `roaming' mechanism for molecular dissociation reactions in terms of
geometrical
structures in phase space. These are NHIMs (Normally Hyperbolic Invariant Manifolds) and their stable/unstable manifolds that define
transition
states for ion-molecule association or dissociation reactions. The associated dividing surfaces rigorously define a roaming region of phase
space,
in which both reactive and nonreactive trajectories can be trapped for arbitrarily long times.
\end{abstract}


\begin{keyword} 
Roaming reactions, transition state theory, phase space structure
\end{keyword}

\end{frontmatter}

\section{Introduction}

Until recently, it was believed that unimolecular dissociations \cite{Levine09}
can occur in either of two ways: (\textit{i}) passage over a potential
energy barrier or (\textit{ii}) barrierless dissociation (e.g., bond fission)
\cite{Bowman2011Suits}.
However, recently a variety of so-called `non-MEP' (Minimum Energy Path)
reactions have been
recognized \cite{Carpenter04,Lourderaj09};
for non-MEP reactions a `reaction coordinate'
cannot be defined in the usual way \cite{Heidrich95}, and statistical theories such as transition state theory
(TST) are not necessarily applicable (see below).

Of particular interest here is the class of {\it ``roaming reactions"} \cite{Bowman2011Suits}.
The roaming phenomenon was discovered in the photodissociation of the formaldehyde
molecule, H$_2$CO \cite{townsend2004roaming}.
In this process, H$_2$CO can dissociate via two channels: H$_2$CO $\rightarrow$ H + HCO (radical channel) and
H$_2$CO $\rightarrow$ H$_2$ + CO (molecular channel).
Above the threshold for the
H + HCO dissociation channel, the CO rotational state distribution
was found to exhibit an intriguing `shoulder' at lower
rotational levels correlated with a hot vibrational distribution of H$_2$ co-product \cite{zee:1664}.
The observed product state distribution did
not fit well with the traditional picture of the
dissociation of formaldehyde via a well characterized
saddle point transition state  for the molecular channel.
The roaming mechanism, which explains
the observations of ref.\cite{zee:1664}, was demonstrated both
experimentally and in trajectory
simulations in ref.\cite{townsend2004roaming}.
Following this work, roaming  has been identified in the
unimolecular dissociation of a number of molecules,
and is now recognized as a general phenomenon in unimolecular decomposition
(see \cite{bowman2011roaming} and references therein).
See also refs \cite{Ulusoy13,grubb12,Yu11,hause11}.

These studies have highlighted some general characteristics 
which a dissocating molecule should have in order to manifest roaming:
the existence of 
competing dissociation channels, such as molecular and radical products;
the existence of a saddle on the
potential energy surface (PES) just below the dissociation threshold for radical production;
long range attraction  between fragments.

Reactions exhibiting roaming
do not fit into conventional  reaction mechanistic schemes,
which are based on  the concept of the reaction coordinate \cite{Heidrich95},
for example, the {\it intrinsic reaction coordinate} (IRC).
The IRC is a MEP in configuration space
that smoothly connects reactants to products, and
according to conventional wisdom it is the path
a system follows as
reaction occurs.
Roaming reactions, instead, avoid the IRC
and involve more complicated dynamical behavior.
The roaming phenomenon seems to arise in the presence of long range
interactions between dissociating fragments,
where the possibility of orientational dynamics of
the two fragments can lead to a different set of products and/or
energy distribution than the one expected from MEP intuition.
However, despite much work, it is still unclear
how general the roaming phenomenon is and, specifically, which classes
of reaction might show similar behaviors.
The unusual nature of roaming reactions is a challenge for
TST, where the aim is to
compute reaction rates for a specific (given) reaction pathway.

TST can take  various forms, such as
RRKM (for Rice, Ramsperger, Kassel, and Marcus) theory \cite{Forst03} or
variational transition state theory (VTST) \cite{Truhlar1984}.
The central ingredient of TST is the concept of a {\it dividing surface} (DS),
which is a surface
the system must cross in order to pass from reactants to products (or the reverse).
Association of transition states with
saddle points on PES (and their vicinity) has a long history of successful
applications in chemistry, and has provided
great insight into reaction dynamics \cite{Levine09,Carpenter84,Wales03}.
Accordingly, much effort has been devoted to connecting roaming reaction pathways
with the existence of particular saddle points on the PES  \cite{shepler2011roaming}.
Nevertheless, it is fair to say that these efforts have not been fully successful
thus far.

Reactions proceeding without a clear correlation to features
on the potential energy surface are likely mediated by transition states that are dynamical in nature,
i.e.  {\it phase space structures}.  The phase space formulation of TST has been
known since the beginning of the theory \cite{Wigner38}.
Only in recent years, however, has the phase space
formulation of TST
reached conceptual and computational maturity \cite{Wiggins08}.
Fundamental to this development is the recognition of the role of phase space objects,
namely {\it normally hyperbolic invariant manifolds} (NHIMs) \cite{Wiggins_book1994},
in the construction of relevant DS for chemical reactions.
While the NHIM approach to TST has enabled a deeper understanding
of reaction dynamics for systems with many ($ \ge 3$) degrees of freedom (DoF)
\cite{Wiggins08,ezra2009microcanonical}, its practical
implementation has relied strongly on mathematical techniques
to compute NHIMs such as normal form theory \cite{Wiggins_book03}.
Normal form theory, as applied to reaction rate theory,
requires the existence of a
saddle of index  $\geq 1$ \cite{Wiggins08} on the PES to construct NHIMs and their attached DSs.
For dynamical systems with 2 DoF the NHIMs are just unstable periodic orbits, which have long been
known in this context as Periodic Orbit Dividing Surfaces (PODS)
(for a review, see ref.\ \cite{Pechukas81}).
As we shall see,
these particular hyperbolic invariant
phase space structures (POs/PODS) are appropriate for describing reaction dynamics in
situations where there is no critical point
of the potential energy surface in the relevant region of configuration space.

As noted, the roaming effect manifests itself in
systems having long range interactions between the two fragments
of a unimolecular decomposition, thus allowing mutual
reorientation dynamics.
Ion-molecule reactions are good candidates
to exhibit the roaming effect, as it is well known that long range interactions determine the dynamics
of this type of reactions in the absence of saddle points on the PES along the MEP \cite{Levine09}.

There has been much debate
concerning  the interpretation of experimental results on ion-molecule reactions \cite{Chesnavich82}.
Some results support a model for reactions taking place via the so-called
loose  or {\it orbiting transition states} (OTS),
while others suggest that the reaction operates
through a {\it tight transition state} (TTS)
(for a review, see ref.\ \cite{Chesnavich82}).
To account for this puzzling situation the concept of
{\it transition state switching} was developed \cite{Chesnavich82},
where both kinds of TS (TTS and OTS) are present and
determine the overall reaction rate.
(See also ref.\cite{Miller76a}.)
Chesnavich presented a simple model to illustrate these ideas  \cite{Chesnavich1986}.

In this work we revisit the Chesnavich model \cite{Chesnavich1986}
in the light of recent developments in TST.
For barrierless systems such as ion-molecule reactions, the concepts of OTS and TTS
can be clearly formulated in terms of well defined
phase space geometrical objects. (For work on
the phase space description of OTS, see refs \cite{child1986mp,Wiesenfeld05}.)
The first goal of the present article is the identification of
these notions with well defined {\it phase space dividing surfaces} attached to
NHIMs.
The second and main goal is an elucidation of the roaming phenomenon
in the context of the
Chesnavich model. The associated potential function,
possessing many features associated with a realistic molecular PES,
leads to dynamics which clearly reveal
the origins of the roaming effect.
Based on our trajectory simulations, we show how the identification of the TTS and OTS
with periodic orbit dividing surfaces (PODS)
provides the natural framework for analysis of the roaming mechanism.

\section{Chesnavich model Hamiltonian}

The transition state switching model was proposed
to account for the competition between multiple transition states in ion-molecule reactions.
Multiple transition states were studied by Chesnavich in
the reaction CH$_4^+$ $\rightarrow$ CH$_3^+$ + H
using a simple model Hamiltonian \cite{Chesnavich1986}.
The model system consists of two parts: a rigid,
symmetric top representing the CH$_3^+$ cation, and a mobile H atom.
We employ Chesnavich's model restricted to
two dimensions (2D) to study roaming.  

The Hamiltonian for zero overall angular momentum is:
\begin{equation}
\label{eq:ham}
H = \frac{p_r^2}{2 \mu} +
\frac{p_{\theta}^2}{2} \left (\frac{1}{I_{CH_3}} + \frac{1}{ \mu r^2} \right ) + V(r,\theta),
\end{equation}
where $r$ is the coordinate giving the distance between the centre
of mass of the CH$_3^+$ fragment and the hydrogen
atom. The coordinate $\theta$ describes the relative orientation of the two fragments,
CH$_3^+$ and H, in a  plane. The momenta
conjugate to these coordinates are $p_r$ and $p_{\theta}$, respectively,
while $\mu$ is the reduced mass of the system and $I_{CH_3}$ is the moment
of inertia of the CH$_3^+$ fragment.

The potential $V(r, \theta)$ describes the so-called transitional mode.
It is generally assumed
that in ion-molecule reactions the different modes of the system separate
into intramolecular (or conserved) and intermolecular
(or transitional) modes.
The potential $V(r,\theta)$ is made up of two terms:
\begin{equation}
\label{eq:pot}
V(r,\theta) = V_{CH}(r)+V_{coup}(r,\theta),
\end{equation}
with:
\begin{subequations}
\label{eq:subpot}
\begin{align}
V_{CH}(r) &= \frac{D_e}{c_1-6} \left\{ 2(3-c_2)\exp
\left[c_1(1-x)\right]
\right.  \nonumber \\     
&  \left.                 
- (4c_2-c_1c_2+c_1) x^{-6} - (c_1-6)c_2 x^{-4} \right\}, \\
V_{coup} (r,\theta) & = \frac{V_0(r)}{2} \left [ 1-\cos(2\theta) \right ], \\
V_0(r) & = V_e \exp \left [-\alpha(r-r_e)^2 \right ].
\end{align}
\end{subequations}
Here $x=r/r_e$, and the parameters for potential $V(r, \theta)$, eq.\ \eqref{eq:pot},
fitted to reproduce data from CH$_4^+$ species are: 
dissociation energy $D_e = 47$ kcal/mol; 
equilibrium distance $r_e = 1.1$ \AA.  Parameters
$c_1 = 7.37$,  $c_2 = 1.61$, fit the polarizability of the 
H atom and yield a stretch harmonic frequency of 3000 cm$^{-1}$.
$V_e = 55$ kcal/mol is the equilibrium barrier height for internal rotation, chosen so that at
$r = r_e$ the hindered rotor has, in the low energy harmonic oscillator limit, 
a bending frequency of 1300 cm$^{-1}$.
The parameter $\alpha$ controls the rate of conversion of the transitional mode from 
angular to radial mode. By adjusting this parameter one can control 
whether the conversion occurs `early' or `late' 
along the reaction coordinate $r$. 
For our study we fix $\alpha = 1$ \AA$^{-2}$, which corresponds to 
a late conversion. 
The masses are taken to be $m_H = 1.007825$ u, $m_C = 12.0$ u, 
and the moment of inertia $I_{CH_3} = 2.373409 $ u\AA$^2$.
A contour plot of the PES $V(r, \theta)$ is shown in Fig.~\ref{fig1}.

\begin{table}[H]
\begin{center}
\begin{tabular}{||c|c|c|c|c||}
\hline\hline
Energy (kcal mol$^{-1}$) &  $r$ (\AA)   &  $\theta$ (radians) & Stability & Label \\
\hline\hline                
-47      & 1.1    &  0            & CC     & EP1  \\
    8      & 1.1    & $\pi/2$ & CS     & EP2 \\
 -0.63  & 3.45   & $\pi/2$  & CS      & EP3  \\
  22.27 & 1.62   & $\pi/2$  & SS      & EP4 \\
\hline\hline
\end{tabular}
\end{center}
\caption{\label{table:equil} Equilibrium points for potential $V(r, \theta)$ ($\alpha = 1$). 
(CC) means a center-center equilibrium point (EP), (CS) a center-saddle EP and (SS) a saddle-saddle EP.}
\end{table}

In Table~\ref{table:equil},
the stationary points of the potential function are tabulated and are labelled
according to their stability. The minimum for CH$_4^+$ (EP1) is of center-center stability type (CC),
which means that it is stable in both coordinates, $r$ and $\theta$.
The saddle, which separates two symmetric minima, at $\theta= 0$ and $\pi$ (EP2), is of center-saddle type (CS),
i.e. stable in $r$ coordinate and unstable in $\theta$. The maximum in the PES (EP4) is a saddle-saddle
equilibrium point (SS). The outer saddle (EP3) is a CS equilibrium point.

The MEP connecting the minimum EP1 with the
saddle EP2 at $r=1.1$ \AA  \; (see Fig.~\ref{fig1})
describes a reaction involving  `isomerisation' between two symmetric minima.
The MEP for dissociation to radical products (CH$_{3}^{+}$ cation and H atom) follows the
line $\theta=0$ with
$r \rightarrow \infty$ and has no potential barrier. Broad similarities between the
features of the Chesnavich model and molecules for
which the roaming reaction has been
observed can readily be identified.
In the Chesnavich model we recognize two reaction `channels', one leading to a molecular product,
in fact to the same molecule, by passage over an inner TS,
and one to radical products via dissociation.
Moreover, a saddle (EP3) exists just below the dissociation threshold.

\section{Results}

\begin{figure}[H]
\includegraphics[scale=0.7]{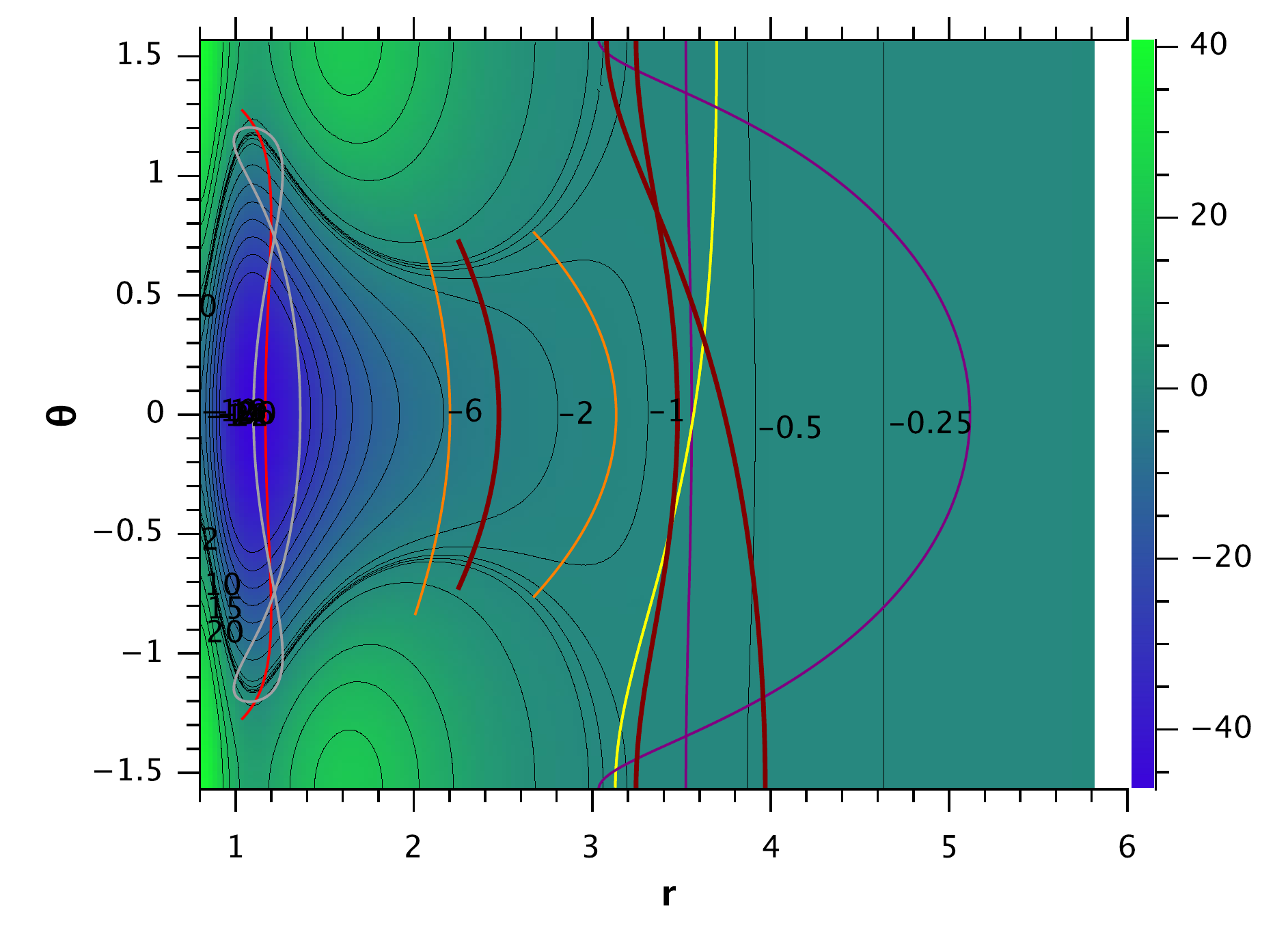}
\caption{\label{fig1} A contour plot of the Chesnavich potential $V(r, \theta)$
(eq.\ \eqref{eq:pot}) with projections of representative periodic orbits.
Orange lines are POs of the $S2th1$ family (definitions of the symbols are given
in Fig. \ref{fig2}),  at energies -0.252 and 4.993 kcal/mol, purple lines are POs of
$S2FR1$ family at energies -0.0602 and 5.005 kcal/mol and yellow line is the period
doubling PO ($S2FR12$) at energy 2.035 kcal/mol.
The wine color POs, all at energy 1.392 kcal/mol, belong from the
left to $S2th1$, $S2FR1$ and $S2FR12$ families, respectively.
The red and grey periodic orbits shown in the region of the
minimum belong to families of the potential which drive the molecule to isomerization.  Distances in Angstroms, angles in radians and
energies in
kcal/mol.}
\end{figure}

Examining the potential in Fig.~\ref{fig1} it is not difficult to
anticipate the existence of two classes of reactive events,
isomerization and direct dissociation to radicals,
but the occurence of a {\it ``third way''} (roaming \cite{Bowman2011Suits}, see below) is
difficult to predict, even for this
simple 2D system. Although it is customary to associate
aspects of a molecule's dynamics with specific features of the PES landscape
\cite{Wales03},
recent progress in non-linear mechanics
suggests caution, especially in the  interpretation
of chemical reactivity.
A method to explore the phase space structure of
a non-linear dynamical system for extended ranges of energy (or other
system parameters) dates back to Poincar\'{e} \cite{poincare1993}, and
involves the study of periodic orbits and their
continuation as energy or other parameters vary.

Distinct  families of POs emanate from equilibrium points, where the
number of families is at least as large as the number of DoF \cite{Weinstein73,Moser76}.
POs of the same family can be followed as energy increases.
At critical values of energy bifurcations take place and
new families are born.
Continuation/bifurcation (CB) diagrams are obtained
by plotting a property of POs as a
function of energy or some other parameter.
One important kind of elementary bifurcation is the center-saddle (CS) (saddle-node)
\cite{Wiggins_book03}.
Although periodic orbits, being  one dimensional objects,  cannot
reveal the full structure of phase space, they do provide
a `skeleton' around which more complex structures such as NHIMs
develop. Numerous explorations of non-linear dynamical systems by
construction of PO CB diagrams have been made
(for molecules, see refs \cite{Farantos98,Farantos09}).

\begin{figure}[H]
\includegraphics[scale=0.7]{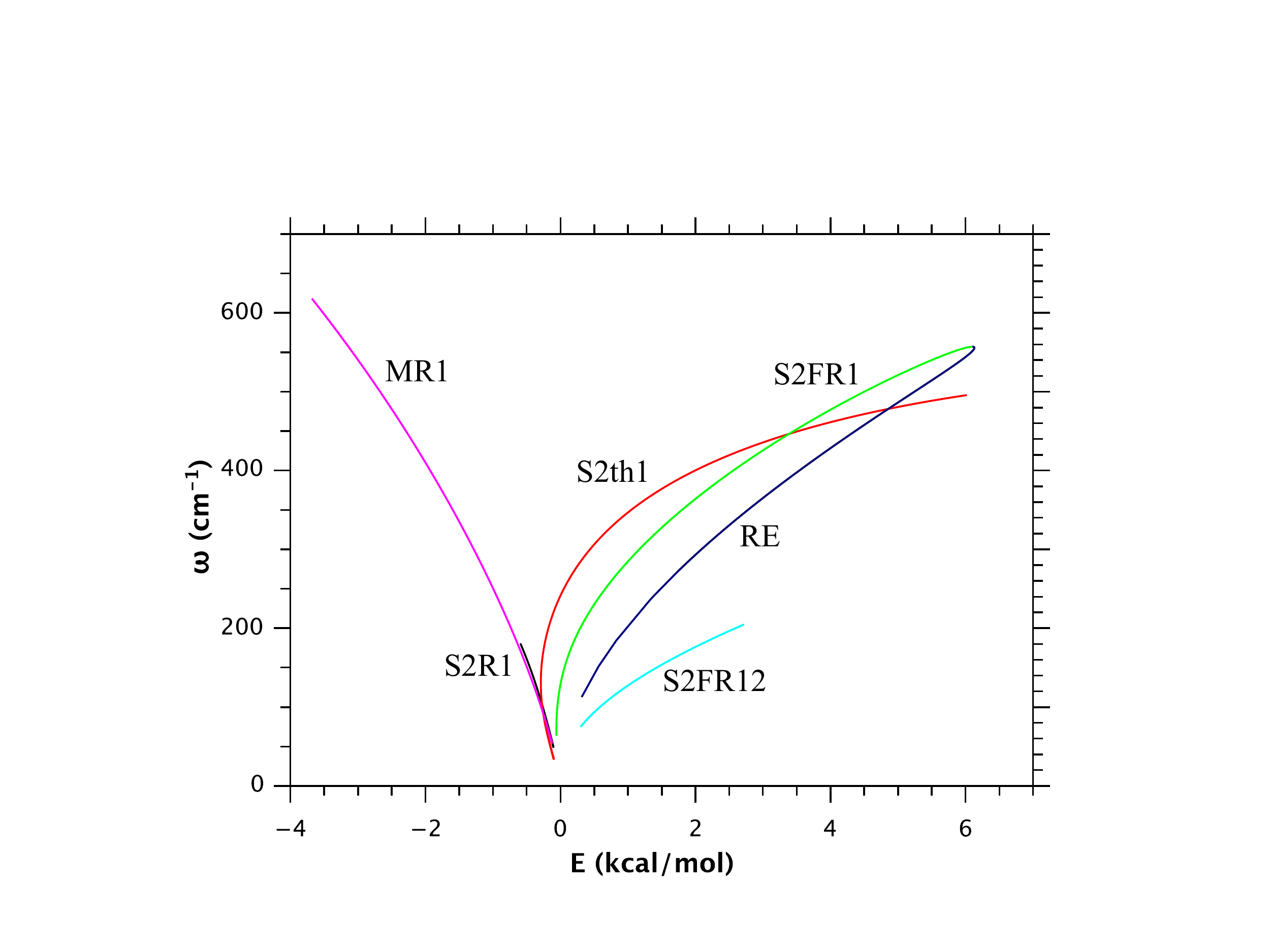}
\caption{\label{fig2} Continuation/Bifurcation diagram of periodic orbits for the
Chesnavich potential and $\alpha=1$. $MR1$ denotes the principal family of
POs along $r$ that originates from the minimum of the potential
(equilibrium point EP1), whereas $S2R1$ is the corresponding family that emanates
from the saddle (EP3). $S2th1$ is the family of POs with hindered rotor behaviour,
acting as the TTS (see text) that emanates from a center-saddle bifurcation and
appears at about $E=-0.291$ kcal/mol below the dissociation energy.
Similarly, $S2FR1$ denotes the family of POs with free rotor behaviour that
also originates from a CS bifurcation at energy $E=-0.0602$ kcal/mol,
while $S2FR12$ a period doubling bifurcation of $S2FR1$ family,
which is generated at energy $E=2.715$ kcal/mol. At this energy the $S2FR1$ becomes unstable.
$RE$ is the family of POs  which are near free rotors and act as the OTS.
They are relative minima,  and have $r \sim \text{const}$ , $p_r=0$ and $p_\theta \ne 0$ also
approximately constant.
This family is the unstable branch of a subcritical CS bifurcation
with the $S2FR1$ family the stable branch, and  emerges at energy 6.131 kcal/mol.}
\end{figure}

In Fig.~\ref{fig2} such a CB diagram is shown
for the Chesnavich model with representative POs depicted in Fig.~\ref{fig1}.
Not all families of POs generated from all equilibria are shown,
but only those which are relevant to the roaming effect.
A detailed description of the various PO families is given in the caption
of Fig.~\ref{fig2}.

The phase space approach to TST requires the identification of NHIMs which serve as
`anchors' for the construction of DSs that locally minimize
the flux. For 2 DoF systems,
the NHIM is just a periodic orbit, which we call the NHIM-PO.
Normal hyperbolicity of
the NHIM-PO implies that  it possesses one stable and one unstable
direction transverse to the PO.
The NHIM-PO is a one dimensional object embedded in the four dimensional phase space.
A dividing surface at a specific energy is a phase space surface
that divides the energy surface into two parts,
namely reactants and products.
The NHIM-PO does not have the right dimensionality
to perform as a DS on the 3 dimensional energy surface in
4 dimensional \emph{phase space}. Rather, the NHIM-PO serves as the
\emph{boundary} of the relevant DS, which is the NHIM-PODS.
The NHIM-PODS at a specific energy is a sphere on which the NHIM-PO is an equator.
The NHIM-PO in turn divides
the NHIM-PODS into two hemispheres, one of
which (the forward hemisphere) intersects all the trajectories which evolve from reactants to products,
while the other (the backward hemisphere)
intersects all the trajectories which travel from products to reactants.

Our first task is then to identify the TTS and the OTS as DSs
attached to appropriate NHIM-POs.
For a system with a natural Hamiltonian (kinetic plus potential terms), when we plot a suitable PO
in the $(r,\theta)$ plane we represent
simultaneously the NHIM-PO and the DS constructed from it.

\begin{figure}[H]
\includegraphics[scale=0.6]{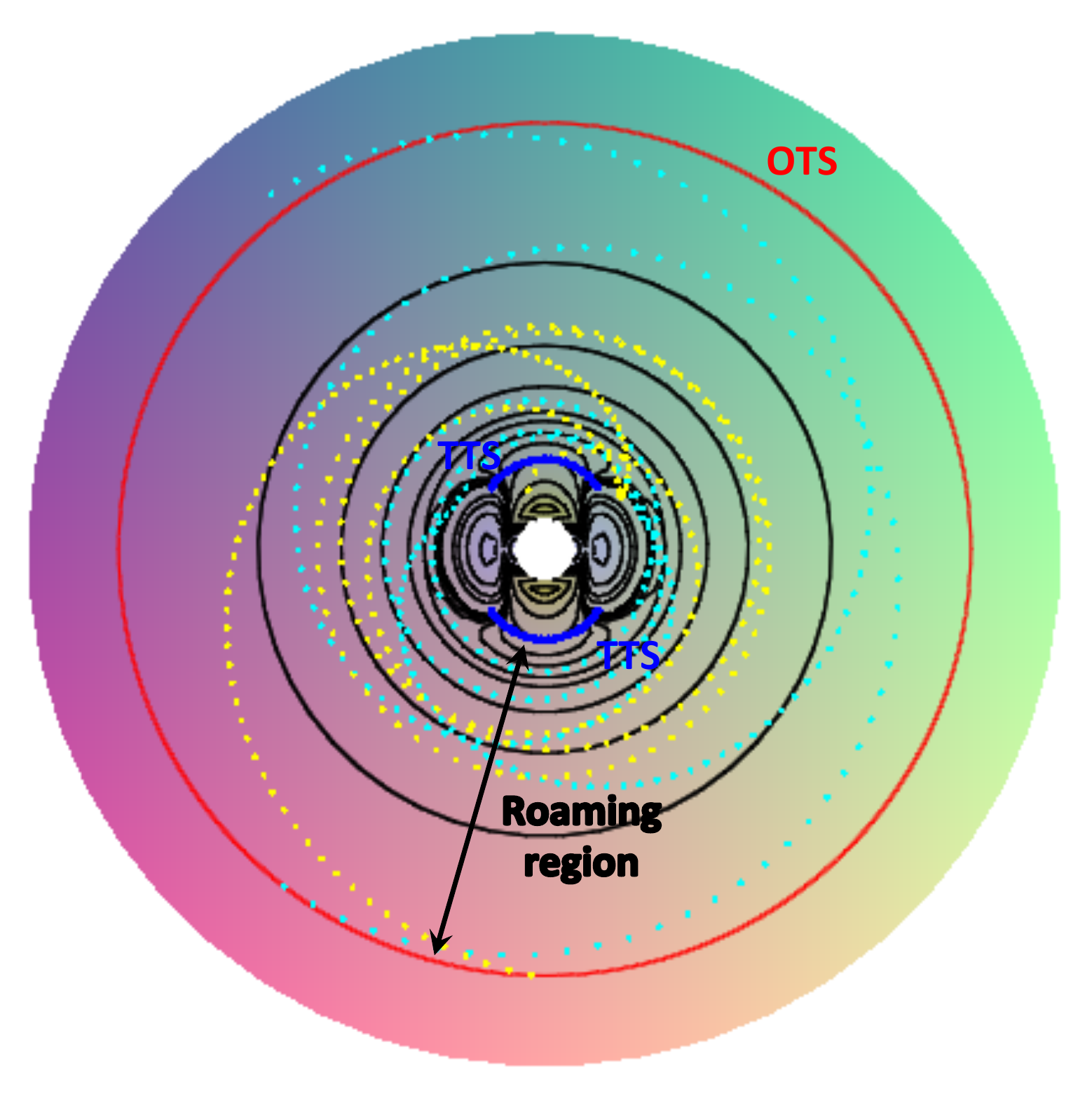}
\caption{\label{fig3} Contour plot of the PES in a Cartesian coordinate system. 
The red line is the projection of OTS and the two blue lines
are the
projections of the TTS. Yellow and
cyan dotted lines represent a roaming reactive and non reactive trajectory respectively (see text).}
\end{figure}

The NHIM-PODS associated with the S2th1 family of periodic orbits
(see  Fig.~\ref{fig2}) are identified with the TTS.
Fig.~\ref{fig3} is a color plot of the potential function in the Cartesian ($xy$) plane
and the two blue lines shown are examples of
two such NHIM-PODS projected on configuration space at an energy corresponding  to a thermal energy of 300 K.
There are two  symmetry-related NHIM-PODS on Fig.~\ref{fig3}.
It has been recognised that the OTS is related to the
centrifugal barrier arising from
the centrifugal term in the kinetic energy, Eq.~(\ref{eq:ham}).
The PO associated with the centrifugal barrier
is a relative equilibrium \cite{Wiesenfeld05}, and this PO belongs
to the RE family shown in Fig.~\ref{fig2}.
These RE POs and higher dimensional analogues
have been studied by Wiesenfeld et \textit{al.}
\cite{Wiesenfeld05} in the context of capture theories of reaction rates.
An example of such a RE NHIM-PODS is depicted as the red outer circle in Fig.~\ref{fig3}
at the thermal energy of 300 K.
We have therefore associated the TTS and OTS with well
defined DSs attached to dynamical objects, i.e. NHIM-PODS.

\begin{figure}[H]
\includegraphics[scale=0.5]{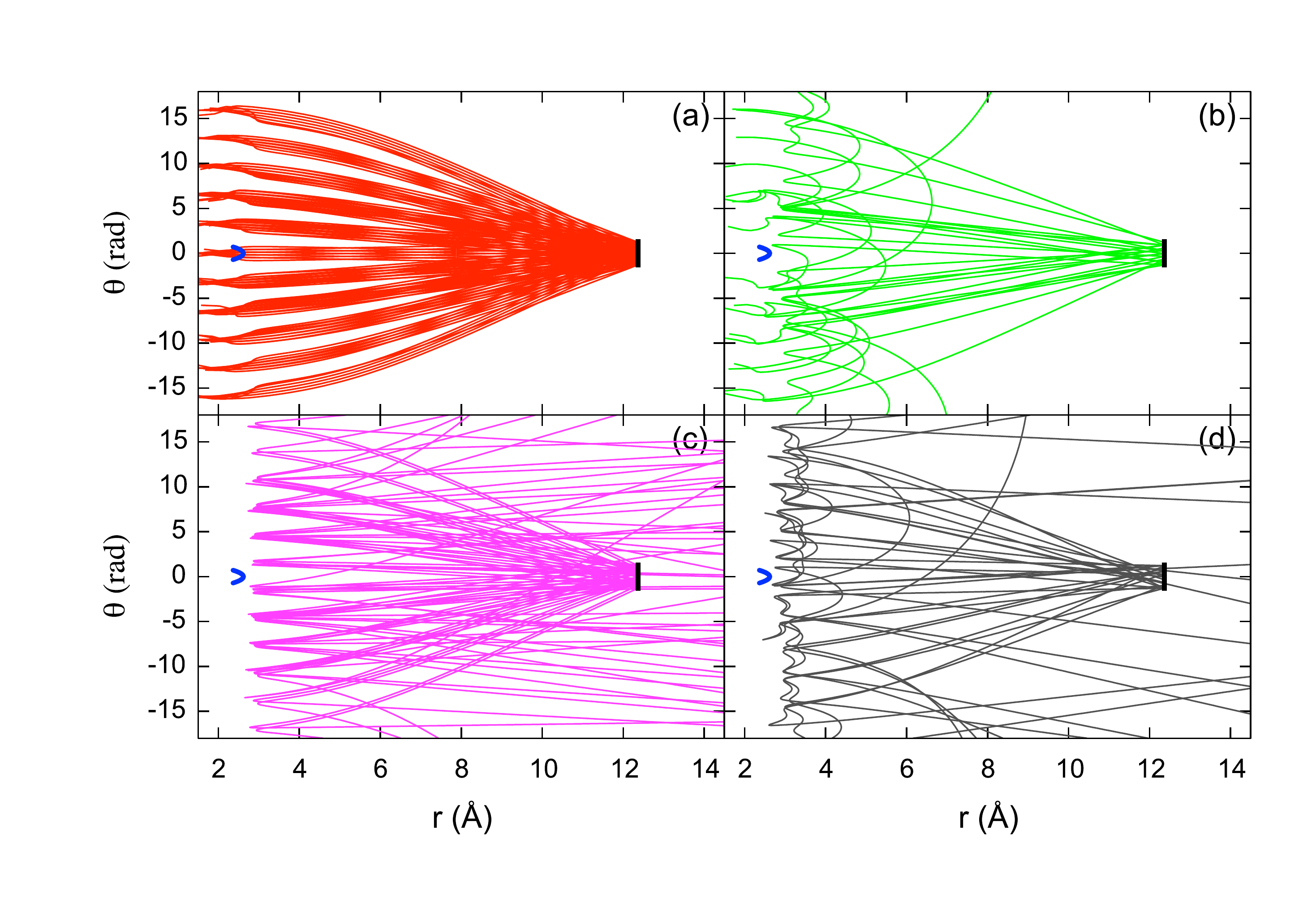}
\caption{\label{fig4} Trajectories initiated on OTS backward hemisphere.
The thick black line represents the OTS and the thick blue line the TTS.
All trajectories have energy $E = k_{\text{B}} T$, where $k_{\text{B}}$ is Boltzmann's
constant and $T = 300$ K.
(a) Direct reactive trajectories. (b) Roaming reactive trajectories.
(c) Direct non reactive trajectories. (d) Roaming non reactive trajectories. }
\end{figure}

To investigate  the behaviour of the trajectories initiated at the OTS, we sample
the DS at the thermal energy 300K.
We sample the backward hemisphere of the DS, which
intersects all trajectories
passing from large values of $r$ into the interaction region (small values of $r$).
The result of this trajectory
simulation is shown in Fig.~\ref{fig4}.
Trajectories are initiated on the black line
segment at $r \sim 12 $ \AA, which is the projection of the OTS on configuration space restricted to the
$\theta$ range $[-\frac{\pi}{2};\frac{\pi}{2}]$.
The DS is sampled uniformly in $\theta$ and the conjugate
variable $p_\theta$ at effectively fixed $r$ and fixed total energy.
For clarity, we do not impose  $\pi$-periodicity  in angle $\theta$
on the plotted trajectories, but rather let this
coordinate increase or decrease freely as the trajectory evolves in time.

We classify the trajectories into
four qualitatively different categories, noting that
a reactive trajectory is one which crosses the (inner) TTS passing from large $r$ to smaller $r$.
(Integration of reactive trajectories terminates shortly after they cross the TTS.)
The four different trajectory categories are:
\begin{itemize}
\item[(\textit{a})] Direct reactive trajectories: these
have no turning points in the $r$ direction, i.e.,
they react directly without making any oscillations in the $r$ direction.

\item[(\textit{b})] Roaming reactive trajectories:
these react but exhibit \emph{at least} two turning points in the $r$ direction.

\item[(\textit{c})] Direct non reactive trajectories:
these trajectories
go to small values of $r$ and are reflected once and recross.

\item[(\textit{d})] Roaming non reactive trajectories:
these trajectories  do not react, but are not direct trajectories.
They never cross the TTS
but eventually recross the OTS to end up at large values of $r$.
However, before recrossing the OTS they exhibit at least  three turning points in $r$.

\end{itemize}
These four categories exhaust all possible qualitatively different trajectory
behaviors (we ignore measure zero sets of trapped trajectories
that approach POs in the roaming region along stable manifolds).
In Fig.~\ref{fig3} in addition to
the TTS and OTS we plot two trajectories,
one roaming reactive (yellow dots) and one roaming non reactive (cyan dots).

With this classification of trajectories the existence of
the roaming phenomenon is immediately
apparent.
Panels (b) and (d) of Fig.~\ref{fig4} show  trajectories which attempt to react
but cannot find their way through
the TTS, and are  reflected back.
Close to the TTS exchange of energy between the radial
and angular modes takes place and the hydrogen atom starts to orbit the CH$_3^+$ ion
in the \emph{roaming region}, which is the region of \emph{phase space}
between the TTS and the OTS,
before perhaps returning and crossing the TTS to react
(panel (b)) or promptly recrossing the OTS and
leaving the interaction region forever (panel (d)).

For the reverse process (photodissociation), we
want to know the behaviour of trajectories initiated on the TTS.
Thus, the trajectories in panel (b)
of Fig.~\ref{fig4} can be thought of as those trajectories
which start at the TTS, roam and then cross the OTS to give CH$_3^+$+H.
Again, the roaming mechanism finds a natural explanation once
we identify the
relevant transition states, i.e. the TTS and the OTS.
These two DS create a trapping region between them,
in which some trajectories may be captured circling for arbitrarily
long times.

\begin{figure}[H]
\includegraphics[scale=0.5]{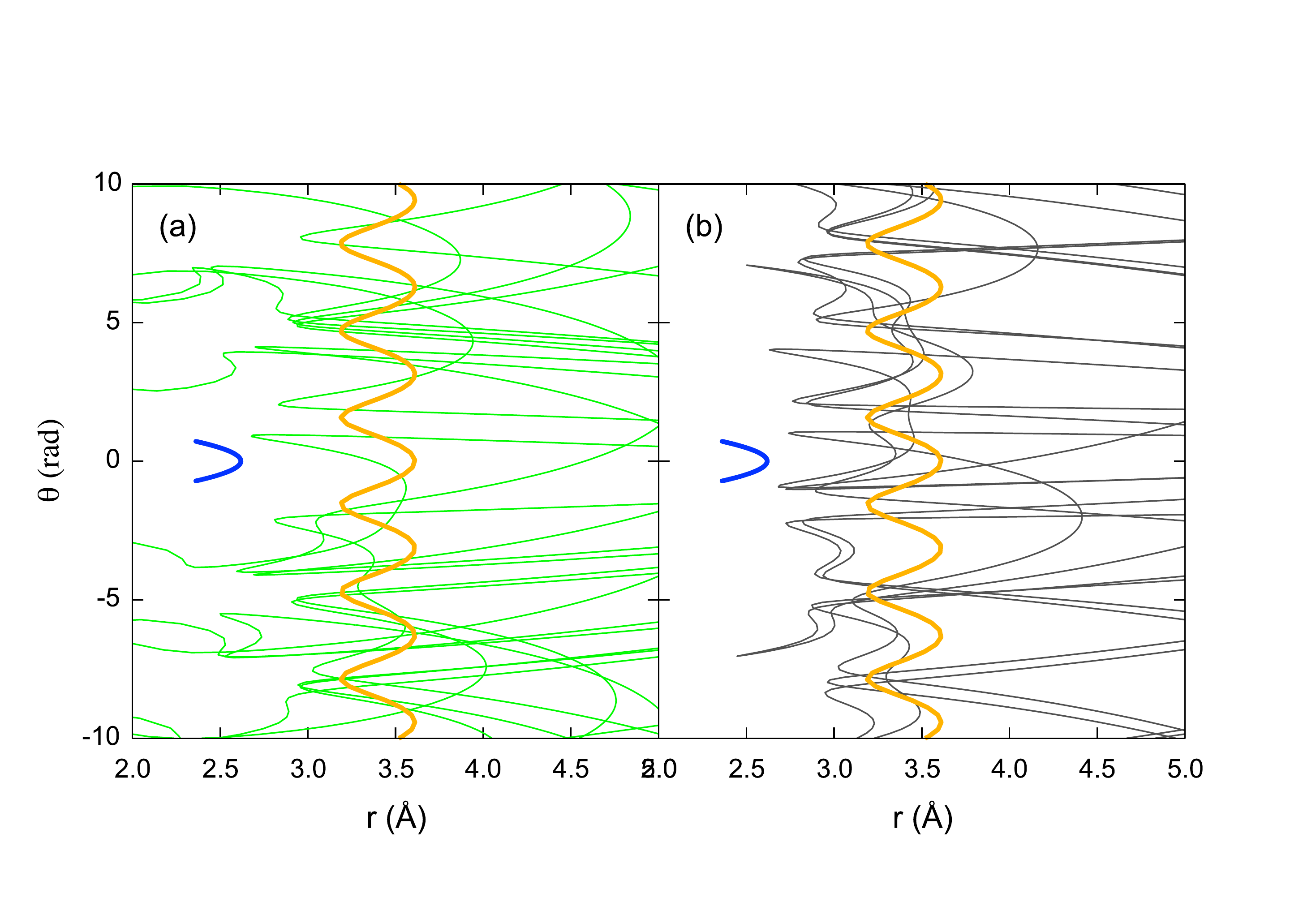}
\caption{\label{fig5}
Trajectories of panels (b) and (d) of Fig.~\ref{fig4} with the PO ($S2FR1$) responsible for
trapping of these trajectories (orange line).}
\end{figure}

From Fig.~\ref{fig4} panels (b) and (d), we can see that trajectories
appear to oscillate in the $r$ direction at $r \sim  3.5$ \AA.
This fact can be explained by the presence of the S2FR1 family
of periodic orbits (see  Figs~\ref{fig1} and \ref{fig2}), where the H atom makes
full rotations in the angle $\theta$ and small oscillations in $r$.
In Fig.~\ref{fig5} we plot the same trajectories as in
panels (b) and (d) of Fig.~\ref{fig4} with the projection of this PO (orange line) for the same energy (300 K).
We see this PO is actually a 2:1 resonance between the
radial and angular modes, since during the time $\theta$ covers the range $[0;2\pi]$
there are two oscillations in the $r$ direction.
Trajectories are presumably trapped by the stable and unstable invariant manifold
of the S2FR1 PO (and/or POs created by period-doubling bifurcations, such as POs of family S2FR12 in Fig.~\ref{fig2}),
which explains the resemblance of the trajectories to this PO.

We note that the S2FR1 family originates at an energy below the threshold
energy for dissociation to radical products, whereas the RE family exists
only for positive energies (cf.\ ref \cite{suits2008}).
We emphasize that, despite the
existence of the saddle EP3, the transition state
that controls the dissociation (association) reaction,
and especially roaming, is that related to the RE periodic orbit.
Calculation of action integrals for the various periodic orbits
shows that for the RE family the action is smaller than S2FR1 POs.
The minimum flux criterion required in TST \cite{Pechukas81} is thus satisfied by the RE POs.
Periodic orbits of S2FR1 type and its period doubling bifurcations, which emerge from above
the saddle EP3, presumably serve to enhance the roaming effect by
increasing the possibilities for trapping of trajectories.

\begin{figure}[H]
\includegraphics[scale=0.5]{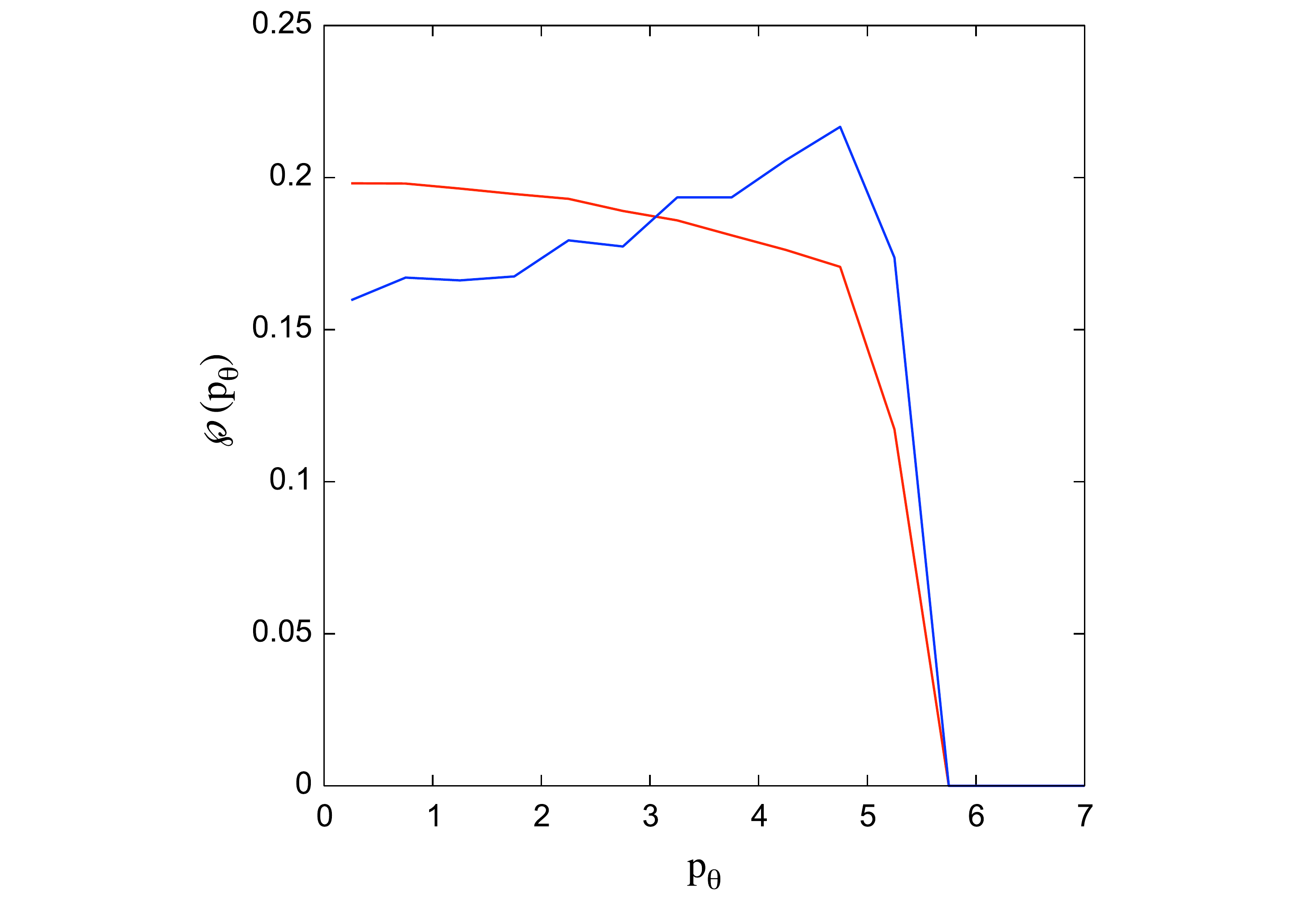}
\caption{\label{fig6}
Rotational state distributions for non-reactive trajectories initiated on the OTS.
Normalized final rotor angular momentum ($p_\theta$) distributions for roaming (blue)
and direct (red) non-reactive trajectories are shown.}
\end{figure}

Direct and roaming
non reactive trajectories exhibit different final rotational state distributions
(see Fig.\ \ref{fig6}), in line with previous findings on the roaming mechanism.
Direct non reactive trajectories are more likely to
suffer a collision with the inner wall of the potential and to
exit the roaming region with large radial kinetic energy, and low
final rotor angular momentum.  Conversely,
trapped (roaming) trajectories are likely to have lower radial kinetic energy and hence
larger rotor angular momentum.

\section{Discussion}

The roaming phenomenon has stimulated much recent research \cite{bowman2011roaming}
and has led to the identification of  the roaming `mechanism' in the dissociation
dynamics of several polyatomic molecules.
For example, Klippenstein et al. \cite{Klippenstein2011} 
have produced a statistical theory for the effect
of roaming pathways on product branching fractions in both unimolecular and bimolecular reactions.
This theory uses approximate dividing surfaces in
configuration space.
A quantum mechanical investigation of the roaming effect 
for the H + MgH $\rightarrow$ Mg + H$_2$ reaction 
at low collision energies
has recently been published \cite{Guo2013jpcA}.

The roaming effect has brought transition state theory once more to
the frontiers of research in chemical dynamics.
In spite of much recent progress in the development of the phase space approach to
fundamental concepts related to TST, such as dividing surface, activated complex and reaction pathways
\cite{Wiggins08},
adherence to a configuration space viewpoint
based on the potential energy surface alone
continues to lead to conceptual difficulties and
confusion when
treating chemical reactions not directly associated with
minimum energy paths or saddles on the PES.

In this article, we have clearly demonstrated that NHIMs
and their stable/unstable manifolds
exist and define transition states for ion-molecule association or dissociation reactions.
The associated DS rigorously define a roaming region of phase space,
in which both reactive and nonreactive trajectories can be trapped for arbitrarily long times.
Extension of the concepts developed here to higher dimensional ($n \geq 3$ degrees of freedom) systems
is in principle straightforward, as our framework does not depend essentially on
dimensionality.  Nevertheless, substantial technical difficulties need to be overcome
for  accurate computation of NHIM-DS for higher dimensional systems.
As new experimental techniques are invented
and allow ever more detailed exploration of molecular phase space,
new exotic reaction pathways will no doubt be discovered and,
perhaps, controlled.

\section*{Acknowledgments}

This work is supported by the National Science Foundation under Grant No.\ CHE-1223754 (to GSE).
FM, PC, and SW  acknowledge the support of the  Office of Naval Research (Grant No.~N00014-01-1-0769),  
the Leverhulme Trust, and the
Engineering and Physical Sciences Research Council (Grant No.~ EP/K000489/1).




\def\cprime{$'$}

\newpage

\newpage

\newpage

\newpage

\newpage

\newpage

\end{document}